\documentclass[onecolumn]{aastex63}

\usepackage{CJK}
\usepackage{amsmath}
\usepackage{multirow}
\usepackage{graphicx}  
\usepackage{float} 
\usepackage{subfigure} 
\usepackage{threeparttable}
\usepackage{captcont}
\usepackage{xcolor}
\usepackage{booktabs}
\usepackage{soul}

\begin{document}

\newcommand*{\mkblue}[1]{\textbf{\color{blue}{#1}}}

\newcommand{\PMO}{Key Laboratory of Dark Matter and Space Astronomy, Purple Mountain Observatory, Chinese Academy of Sciences, Nanjing 210033, People's Republic of China.}
\newcommand{\USTC}{School of Astronomy and Space Science, University of Science and Technology of China, Hefei, Anhui 230026, People's Republic of China.}
\newcommand{\AHNU}{Department of Physics, Anhui Normal University, Wuhu, Anhui 241000, People's Republic of China.}
\newcommand{\GXU}{Guangxi Key Laboratory for Relativistic Astrophysics, Nanning 530004, People's Republic of China.}

\title{Potential Subpopulations and Assembling Tendency of the Merging Black Holes}

\author{Yuan-Zhu Wang}
\affiliation{\PMO}
\author{Yin-Jie Li}
\affiliation{\PMO}
\affiliation{\USTC}
\author{Jorick S. Vink}
\affiliation{Armagh Observatory and Planetarium}
\author{Yi-Zhong Fan}
\author{Shao-Peng Tang}
\affiliation{\PMO}
\author{Ying Qin}
\affiliation{\AHNU}
\affiliation{\GXU}
\author{Da-Ming Wei}
\affiliation{\PMO}
\affiliation{\USTC}

\correspondingauthor{Yi-Zhong\, Fan; Ying\, Qin; Jorick S. Vink}
\email{yzfan@pmo.ac.cn; yingqin2013@hotmail.com; Jorick.Vink@Armagh.ac.uk}

\date{\today}

\begin{abstract}
The origins of coalescing binary black holes (BBHs) detected by the advanced LIGO/Virgo are still under debate, and clues may be present in the joint mass-spin distribution of these merger events. Here we construct phenomenological models containing two sub-populations to investigate the BBH population detected in gravitational wave observations.
We find that our models can explain the GWTC-3 data rather well, and several constraints to our model are required by the data: first, the maximum mass for the component with a stellar-origin, $m_{\rm max}$, is $39.1^{+2.4}_{-2.7}M_{\odot}$ at 90\% credibility; second, about $15\%$ of the mergers happen in dynamical environments, in which $7-16\%$ of events are hierarchical mergers, and these BHs have an average spin magnitude significantly larger than the first-generation mergers, with ${\rm d}\mu_{\rm a} > 0.4 $ at $99\%$ credibility; third, the dynamical component BHs tend to pair with each other with larger total mass and higher mass ratio. An independent analysis focusing on spins is also carried out, and we find that the spin amplitude of component BHs can be divided into two groups according to a division mass $m_{\rm d} = 46.1^{+5.6}_{-5.1}M_{\odot}$. These constraints can be naturally explained by current formation channels, and our results suggest that some of the observed events were likely from AGN disks.
\end{abstract}

\section{Introduction}\label{sec:intro}
With the help of the Advanced LIGO and Virgo detector network, the number of observed binary black hole (BBH) mergers is rapidly growing. To date, more than 90 BBH merger candidates have been released with the current catalogs of compact binary coalescences (GWTC-3) \citep{abbottO2,abbottO3a,2021arXiv211103606T}.
The origins of these compact objects are still unclear. Various formation channels have been proposed, including, for instance, isolated binary evolution, dynamical capture, and AGN enhancement (see~\citealt{mapelli2021review} and \citealt{2021NatAs...5..749G} for recent reviews). 

Formation channels may leave their `fingerprints' in the final population of BHs. One scenario for the origin of BBHs involves the evolution of field binaries. The remnants after the death of very massive stars are expected to fall outside a gap starting at $\sim 40-65 M_\odot$ and ending at $\sim 125M_\odot$ due to (pulsational) pair-instability supernovae ((P)PISNe) \citep{2016A&A...594A..97B,2017ApJ...836..244W,2017MNRAS.470.4739S,2019ApJ...882..121S,2020ApJ...888...76M}. The BH spins are generally small \citep{2018A&A...616A..28Q,2019MNRAS.485.3661F,2020A&A...636A.104B} under the assumption of efficient angular momentum transport inside their progenitors. On the other hand, in the chain of hierarchical mergers, BHs inside the (P)PISNe mass gap can appear in second or higher-generation mergers, and large spin magnitudes are expected \citep{miller2002,giersz2015,fishbach2017,gerosa2017,rodriguez2019,2020ApJ...893...35D,2020ApJ...900..177K,arcasedda2021b,mapelli2021,2021NatAs...5..749G}. More specifically, for hierarchical mergers happening inside AGN disks, BHs may align their orbits and component spins with the disk, leading to a broad effective spin ($\chi_{\rm eff}$) distribution with positive mean value \citep{2019PhRvL.123r1101Y,2021MNRAS.507.3362T}. Please note that the formation channels discussed above involve conventional concepts, and there are still many theoretical uncertainties that may affect the properties of the remnant BHs (see Sec.~\ref{sec:astro} for more discussions).

Building phenomenological models to describe the population properties of BBHs is a widely used approach to reveal the key features that could shed light on the formation channels. To capture the main features of the mass distribution for BBHs with a stellar origin, \citet{2018ApJ...856..173T} proposed an astrophysically motivated model consisting of a power-law component and a Gaussian component. A cutoff mass ($m_{\rm max}$) for the power law is taken as a parameter to represent the lower edge of the PISN mass gap, and the Gaussian peak is an excess due to PPISNe. 
While a peak around $35 M_\odot$ in the primary mass distribution was revealed by the recent study of \citet{2021arXiv211103634T} using the latest catalog (see also the in the analysis of some non-parametric approaches\citep{2021ApJ...913L..19T,2021ApJ...917...33L}), evidence for an upper mass gap was absent. One solution to this discrepancy is that the observed events may originate from multiple channels \citep{2021ApJ...915L..35K,2021ApJ...921L..15G,2021ApJ...913...42W,bouffanais2021,zevin2021,wong2021,roulet2021,2022ApJ...933L..14L}, thus the mass gap is filled by hierarchical events; finding the exact position of $m_{\rm max}$ remains an important goal in the population studies \citep{2017ApJ...851L..25F,2021ApJ...913...42W,2021ApJ...913L..23E,2021ApJ...916L..16B}. 
The picture for the spin distribution is even more ambiguous, as it is harder to measure the component spins from the GW signals. In the analysis of GWTC-2 events by \citet{2021ApJ...913L...7A}, the results for the Default spin model \citep{abbottO2popandrate} and the Gaussian spin model \citep{2020ApJ...895..128M} indicated the presence of extremely misaligned spin. However, \citet{roulet2021} argued that if the model allows for a sub-population with negligible spin, the evidence for negative spins diminishes. Similar results are also obtained by \citet{2021ApJ...921L..15G}, who concluded that a model consisting of a non-spinning sub-population and a rapidly spinning nearly aligned sub-population could better explain the data with respect to LVKC's Default spin model. A recent population study on GWTC-3 by \citet{2021arXiv211103634T} showed that when the non-spinning sub-population is considered, the data still prefer a negative minimum for the distribution of effective inspiral spins. Very recently, both \citet{2022ApJ...937L..13C} and \citet{2022arXiv220902206T} suggested that there is no evidence for the existence of a non-spinning sub-population, and they found support for the presence of mergers with extreme spin tilt angles. Nevertheless, \citet{2022arXiv220906978V} found that the exact shape for the inferred distribution of tilt angle of component BHs strongly depends on the assumed population model and the priors for the model parameters.

The investigations mentioned above may imply that the results obtained from phenomenological models face certain degrees of model dependency due to the finite flexibility of the models. In this work, we carry out the population study from another angle: we assume that the BBHs consist of a ``field" sub-population and a ``dynamical" sub-population, and incorporate the pairing functions \citep{2020ApJ...891L..27F} into the construction of our model. We seek the constraints on the key features and branch ratios of these two components from GWTC-3 events and discuss the requirements on the astrophysical conditions to produce such sub-populations. We also perform a spin-dominant analysis to investigate if there is a correlation between the spins and masses of component BHs to validate the picture of our model. The rest of the paper is arranged as follows: in Sec.~\ref{sec:model}, we describe our models; in Sec.~\ref{sec:method}, we introduce the statistical method; the analysis results are presented in Sec.~\ref{sec:result}; Sec.~\ref{sec:spin} is our spin-dominant analysis, and we give further astrophysical implication and discussions on our results in Sec.~\ref{sec:astro} and Sec.~\ref{sec:summary}, respectively. 

\section{Population Models}\label{sec:model}
One way to model the distribution of the component masses ($m_1,m_2$) of the merging BBHs is to construct the marginal primary mass distribution and the conditional secondary mass distribution \citep{abbottO2popandrate}:
\begin{equation}\label{eq:1}
p(m_1,m_2|\mathbf{\Lambda}_{\rm m})=p(m_1|\mathbf{\Lambda}_{\rm m})p(m_2|m_1,\mathbf{\Lambda}_{\rm m}),
\end{equation}
where $\mathbf{\Lambda}_{\rm m}$ is the hyper-parameters governing the exact shape of the distribution. A form of $p(m_2|m_1,\mathbf{\Lambda}_{\rm m}) \propto m_2^{\beta}$ is generally considered in the literature \citep{abbottO2popandrate,2021ApJ...913L...7A,2021arXiv211103634T} to reflect the dependency between $m_1$ and $m_2$. On the other hand, \citet{2020ApJ...891L..27F} introduced the `pairing function' to study the mass distribution. Following their method, the distribution can be written as
\begin{equation}\label{eq:2}
p(m_1,m_2|\mathbf{\Lambda}_{\rm m_1},\mathbf{\Lambda}_{\rm m_2},\mathbf{\Lambda}_{\rm p}) \propto 
\begin{cases}
p(m_1|\mathbf{\Lambda}_{\rm m_1})p(m_2|\mathbf{\Lambda}_{\rm m_2})w(m_1,m_2|\mathbf{\Lambda}_{\rm p}),&~\text{for}~ m_2<m_1, \\
0, & ~\text{for}~ m_2>m_1.
\end{cases}
\end{equation}
which can be interpreted as follows: the primary and secondary masses are separately drawn from independent distributions $p(m_1|\mathbf{\Lambda}_{\rm m_1})$ and $p(m_2|\mathbf{\Lambda}_{\rm m_2})$, and the probability of two masses belonging to a merging binary is given by the pairing function $w(m_1,m_2|\mathbf{\Lambda}_{\rm p})$ \citep{2020ApJ...891L..27F}. We find it convenient to use the pairing function to address diverse preferences for the combination of component masses in multiple channels, so our following modeling is based on Eq.~(\ref{eq:2}).

We consider the distribution model consisting of two sub-populations: one formed through the evolution of field binaries (labeled with `field'), and the other formed dynamically (labeled with `dyn'). The model that contains both the mass and spin parameters ($\mathbf{\Theta} = (m_1,m_2,a_1,a_2,\cos{\theta_1},\cos{\theta_2)}$) can be expressed as
\begin{equation}\label{eq:3}
p(\mathbf{\Theta}|\mathbf{\Lambda}_{\rm field},\mathbf{\Lambda}_{\rm dyn},f_{\rm dyn}) = (1-f_{\rm dyn}) \ p_{\rm field}(\mathbf{\Theta}|\mathbf{\Lambda}_{\rm field}) + f_{\rm dyn} \ p_{\rm dyn}(\mathbf{\Theta}|\mathbf{\Lambda}_{\rm dyn}),
\end{equation}
where $\mathbf{\Lambda}_{\rm field}$ and $\mathbf{\Lambda}_{\rm dyn}$ are the hyper-parameters for the field and dynamical sub-population respectively, and $f_{\rm dyn}$ is the mixing fraction of the dynamically formed BBHs. Assuming the component BHs in a merging binary are drawn from the same underlying distribution, we extend Eq.~(\ref{eq:2}) to model each sub-population:
\begin{equation}\label{eq:4}
p_{x}(\mathbf{\Theta}|\mathbf{\Lambda}_{x}) \propto  \left [ \prod_{i}^{2} \pi_{x}(m_i,a_i,\cos{\theta_i}|\mathbf{\Lambda}_{x}) \right ] w_x(\mathbf{\Theta}|\mathbf{\Lambda}_{x}) ,
\end{equation}
where $x$ represents `${\rm field}$' or `${\rm dyn}$'. $\pi_x$ is the underlying joint distribution, and we assume that it consists of two independent parts:
\begin{equation}\label{eq:5}
\pi_{x}(m_i,a_i,\cos{\theta_i}|\mathbf{\Lambda}_{x}) = \pi_{{\rm m},x}(m_i|\mathbf{\Lambda}_{{\rm m},x})\pi_{{\rm s},x}(a_i,\cos{\theta_i}|\mathbf{\Lambda}_{{\rm s},x}),
\end{equation}
in which $\pi_{{\rm m},x}$ is the underlying mass distribution and $\pi_{{\rm s},x}$ is the spin distribution for the underlying BHs \textit{if they pair to become a component of merging system.}

For the `${\rm field}$' channel, 
\begin{equation}\label{eq:6}
\pi_{\rm m,field}(m_i|\mathbf{\Lambda}_{\rm field}) = \mathcal{P}(m_i|\alpha, m_{\rm min}, m_{\rm max}, \delta_{\rm m}),
\end{equation}
and,
\begin{equation}\label{eq:7}
\pi_{\rm s,field}(a_i,\cos{\theta_i}|\mathbf{\Lambda}_{\rm field}) = \mathcal{G}(a_i|\mu_{\rm a,field},\sigma_{\rm a,field})\mathcal{G}'(\cos{\theta_i}|\mu_{\rm ct,field},\sigma_{\rm ct,field}),
\end{equation}
where $\mathcal{P}$ is a truncated power-law with a smooth tail at low mass as described in \citet{abbottO2popandrate}; $\mathcal{G}$ and $\mathcal{G}'$ are Gaussian distributions truncated at $[0,1]$ and $[-1,1]$ respectively. Specifically, under the assumptions of efficient angular momentum transport in stellar evolution \citep{2002A&A...381..923S,2019MNRAS.485.3661F}, one would expect $\mu_{\rm a,field} \ll 1$\footnote{Although the second-born BH may spin fast due to tidal acceleration, the vast majority of BHs observed in the horizon of current GW detector networks will have small spins as demonstrated in \cite{bavera2020}.} and $\mu_{\rm ct,field} \sim 1$ \citep{2018A&A...616A..28Q,2019MNRAS.485.3661F,2020A&A...636A.104B} for the binaries formed through the common envelope phase, and the $m_{\rm max}$ should be consistent with the lower edge of PPISN gap \citep{2016A&A...594A..97B}. To approximate the preference for symmetric systems in the binary evolution channel, following \citet{2020ApJ...891L..27F}, we consider a pairing function of
\begin{equation}\label{eq:8}
w_{\rm field}(m_1,m_2|\mathbf{\Lambda}_{\rm field}) \propto \left ( \frac{m_2}{m_1} \right )^{\beta_{\rm field}}. 
\end{equation}
 
For the dynamical channel, building a phenomenological distribution model is more challenging, as the formation of binaries may happen in different environments, and there might be contributions from hierarchical mergers. To avoid our model being too complicated and redundant, we only consider hierarchical mergers up to the second generation. We first consider a relatively simple case in which the first generation (1G) dynamical BBHs share the same underlying mass distribution with those originated from field binaries, i.e., $\pi_{\rm m,1G}(m_i) = \pi_{\rm m,field}(m_i)$. The spin distribution of 1G dynamical BBHs is modeled as
\begin{equation}\label{eq:9}
\pi_{\rm s,1G}(m_i,a_i,\cos{\theta_i}|\mathbf{\Lambda}_{\rm field}) = \mathcal{G}(a_i|\mu_{\rm a,1G},\sigma_{\rm a,1G})\mathcal{G}'(\cos{\theta_i}|\mu_{\rm ct,1G},\sigma_{\rm ct,1G}).
\end{equation}
Assuming the component BHs are the remnants of single star evolution, under the efficient angular momentum transport assumption, their spin magnitudes will also be small as the field BBHs (hereafter, we use `stellar BHs' to represent the field and 1G dynamical BHs); therefore, we let $\mu_{\rm a,1G} \equiv \mu_{\rm a,field} \equiv \mu_{\rm a,ste}$ and $\sigma_{\rm a,1G} \equiv \sigma_{\rm a,field} \equiv \sigma_{\rm a,ste}$ to simplify our model. Guided by the results of some simulations for dynamical processes \citep{oleary2016,yang2019}, we consider a pairing function that depends on both the mass ratio and the total mass:
\begin{equation}\label{eq:10}
w_{\rm dyn}(m_1,m_2|\mathbf{\Lambda}_{\rm dyn}) \propto \left ( \frac{m_2}{m_1} \right )^{\beta_{\rm dyn}}\left ( m_1+m_2
\right )^{\gamma_{\rm dyn}}. 
\end{equation}
In general, dynamical captures in environments such as globular, nuclear, and young star clusters isotropic spin orientation is expected ($\sigma_{\rm ct,dyn} \gg 0$) \citep{2010CQGra..27k4007M,2016ApJ...832L...2R,2022MNRAS.511.5797M}, while the evolution of BBHs in AGN disks may lead to certain degrees of alignment, depending on the details of the dynamics in the disk \citep{mckernan2018,2020MNRAS.494.1203M,2020ApJ...899...26T,2022ApJ...931...82V,2022MNRAS.511.5797M}.

The underlying mass distribution of 2G dynamical BBHs can be approximated by the total mass distribution of 1G mergers, which can be calculated by
\begin{equation}\label{eq:11}
\pi_{\rm m,2G}(m_i|\mathbf{\Lambda}_{\rm m,2G}) \propto \int_{m_{\rm min}}^{m_{\rm max}} \mathrm{d}m_2 \mathcal{P}(m_i^{*}-m_2)\mathcal{P}(m_2)w_{\rm dyn}(m_i^{*}-m_2,m_2)\mathcal{H}(\frac{m_i^{*}}{2}-m_2),
\end{equation}
where $\mathbf{\Lambda}_{\rm m,2G}=(\alpha,m_{\rm min},m_{\rm max},\delta_{\rm m},\beta_{\rm dyn},\gamma_{\rm dyn})$, and we set $m_i^{*} = 1.05m_i$ to approximate the loss of gravitational mass during the coalescence. The underlying joint distribution for 2G black holes can then be written as 
\begin{equation}\label{eq:12}
\pi_{\rm 2G}(m_i,a_i,\cos{\theta_i}|\mathbf{\Lambda}_{\rm 2G}) = \pi_{\rm m,2G}(m_i|\mathbf{\Lambda}_{\rm m,2G})\mathcal{G}(a_i|\mu_{\rm a,2G},\sigma_{\rm a,2G})\mathcal{G}'(\cos{\theta_i}|\mu_{\rm ct,2G},\sigma_{\rm ct,2G}).
\end{equation}
Assuming the dynamical environments contain both 1G and 2G BHs, by introducing a parameter $r_{\rm 2G}$ to describe the fraction of 2G underlying BHs, the overall 1G+2G underlying distribution for the dynamical channel is
\begin{equation}\label{eq:13}
\pi_{\rm dyn}(m_i,a_i,\cos{\theta_i}|\mathbf{\Lambda}_{\rm dyn}) = (1-r_{\rm 2G})\pi_{\rm 1G}(m_i,a_i,\cos{\theta_i}|\mathbf{\Lambda}_{\rm 1G}) + r_{\rm 2G}\pi_{\rm 2G}(m_i,a_i,\cos{\theta_i}|\mathbf{\Lambda}_{\rm 2G}).
\end{equation}

We take the spin tilt hyper-parameters $\mu_{\rm ct,field} \equiv \mu_{\rm ct,1G} \equiv \mu_{\rm ct,2G} \equiv 1$ in our analysis because a half Gaussian that peaks at $1$ and truncates at $-1$ can approximate both the aligned case (with $\sigma \sim 0$) and the isotropic case (with $\sigma \gg 0$). Meanwhile, we also assume that $\sigma_{\rm ct,1G} \equiv \sigma_{\rm ct,2G} \equiv \sigma_{\rm ct,dyn}$ in this work. 

In addition to the fiducial model (namely the OnePL model) described above, we also consider five additional models for comparison: 1) a Gaussian component representing the pile-up of remnants BHs whose progenitors underwent PPISN is added to the underlying mass distributions \citep{2018ApJ...856..173T} of `field' BBHs and 1G `dyn' BBHs (Model PPISN Peak); 2) the 1G `dyn' BHs have a different underlying power-law mass distribution (Model TwoPL); 3) an isotropic spin orientation model for the dynamical channel (Model IsoDyn); 4) the Power-law + Peak mass model and the Default spin model in \citet{2021arXiv211103634T} (Model PP \& Default); 5) having the same mass model as the OnePL model but with the Default spin model (Model OnePL \& Default). We summarise the analytical functions and their parameters of these models in Tab.~\ref{tab:models}, and list the meanings and priors for the parameters of the fiducial model in Tab.~\ref{tab:hp}.

\section{Hierarchical Bayesian Inference}\label{sec:method}
We perform hierarchical Bayesian inference to constrain our model parameters. The likelihood for the inference is constructed based on In-homogeneous Poisson process. For a series of measurements of $N_{\rm obs}$ events $\vec{d}$, assuming a redshift evolving merger rate of $R \propto (1+z)^{2.7}$ (as suggested by \citet{2021arXiv211103634T}), the likelihood for the hyper-parameters $\mathbf{\Lambda}$ can be inferred via \citep{2019PASA...36...10T,2021ApJ...913L...7A}
\begin{equation}\label{eq:llh}
\mathcal{L}(\vec{d}\mid \mathbf{\Lambda}) \propto N^{N_{\rm obs}}\exp(-N \eta(\mathbf{\Lambda}))\prod_{i}^{N_{\rm obs}}\frac{1}{n_i}\sum_{k}^{n_i}\frac{p(\theta_{i}^k\mid \mathbf{\Lambda})}{p(\theta_{i}^k\mid \varnothing)},
\end{equation}
where $N$ is the expected number of mergers during the observation period and can be derived by integrating the merger rate over the co-moving space-time volume. $\eta(\mathbf{\Lambda})$ is the detection efficiency, following the procedures described in \citet{2021ApJ...913L...7A}, we use the injection campaign released in \citet{2021arXiv211103634T} to estimated this quantity. The $n_i$ posterior samples for the $i$-th event and the default prior $\pi(\theta_k \mid \varnothing)$ are obtained from the released data accompanying with \citep{abbottO2,abbottGWTC-2.1,2021arXiv211103606T}. We use the same criteria that define the detectable events as \citet{2021arXiv211103634T}, i.e., ${\rm FAR} < 1/{\rm yr}$, and 69 BBH events passed the threshold cut\footnote{A stricter criterion of ${\rm FAR} < 0.25/{\rm yr}$ are also adopted in the inference for comparison, and it yields consistent results.}. We use the python package {\sc Bilby} \citep{2019ApJS..241...27A} and the {\sc PyMultinest} sampler \citep{2016ascl.soft06005B} to obtain the Bayesian evidence and posteriors of the hyper-parameters for each model.

\begin{table*}[htpb]
\caption{Summary of the functions used in the models and their parameters}\label{tab:models}
\begin{tabular}{c|c|c|c}
\hline
\hline
\multirow{2}{*}{property}  & \multirow{2}{*}{channel} & underlying distribution & \multirow{2}{*}{pairing function} \\
&&(parameters)&\\
\cline{1-4}
\multirow{6}{*}{component mass} & \multirow{2}{*}{field} & smoothed truncated power-law  & \multirow{2}{*}{$\propto \left ( \frac{m_2}{m_1} \right )^{\beta_{\rm field}}$} \\
&&($\alpha_{\rm field},m_{\rm min},m_{\rm max},\delta_{\rm m}$)&\\
\cline{2-4}
& \multirow{2}{*}{1G dynamical} & smoothed truncated power-law  & \multirow{4}{*}{$\propto \left ( \frac{m_2}{m_1} \right )^{\beta_{\rm dyn}}\left ( m_1+m_2
\right )^{\gamma_{\rm dyn}}$} \\
&&($\alpha_{\rm dyn},m_{\rm min},m_{\rm max},\delta_{\rm m}$)&\\
\cline{2-3}
& \multirow{2}{*}{2G dynamical} & calculated from the result &\\
&& of 1G dynamical mergers &\\
\hline
\multirow{6}{*}{spin magnitude} & \multirow{2}{*}{field} & truncated Gaussian  & \multirow{6}{*}{--} \\
&&($\mu_{\rm a,field},\sigma_{\rm a,field}$)&\\
\cline{2-3}
& \multirow{2}{*}{1G dynamical} & truncated Gaussian &\\
&&($\mu_{\rm a,1G},\sigma_{\rm a,1G}$)&\\
\cline{2-3}
& \multirow{2}{*}{2G dynamical} & truncated Gaussian &\\
&& ($\mu_{\rm a,2G},\sigma_{\rm a,2G}$) &\\
\hline
 & \multirow{2}{*}{field} & truncated Gaussian  & \multirow{6}{*}{--} \\
 &&($\mu_{\rm ct,field},\sigma_{\rm ct,field}$)&\\
\cline{2-3}
cosine of spin & \multirow{2}{*}{1G dynamical} & truncated Gaussian &\\
tilted angle &&($\mu_{\rm ct,1G},\sigma_{\rm ct,1G}$)&\\
\cline{2-3}
 & \multirow{2}{*}{2G dynamical} & truncated Gaussian &\\
 && ($\mu_{\rm ct,2G},\sigma_{\rm ct,2G}$) &\\
\hline
\hline
\end{tabular}
\tablenotetext{}{{\bf Note.} (i). For all models, we set $\mu_{\rm a,field} \equiv \mu_{\rm a,1G} \equiv \mu_{\rm a,ste}$, $\sigma_{\rm a,field} \equiv \sigma_{\rm a,1G} \equiv \sigma_{\rm a,ste}$, and $\sigma_{\rm ct,1G} \equiv \sigma_{\rm ct,2G} \equiv \sigma_{\rm ct,dyn}$. We also fix $\mu_{\rm ct,field} \equiv \mu_{\rm ct,1G} \equiv \mu_{\rm ct,2G} \equiv 1$ in the inference; (ii). For the OnePL model, we set $\alpha_{\rm field} \equiv \alpha_{\rm dyn} \equiv \alpha$; The OnePL \& IsoDyn model is the same as OnePL model, except that we fix $\sigma_{\rm ct,dyn}$ to a very large value to represent the case in which the spin of dynamical BBHs tilt isotropically; (iii). see Appendix.~\ref{sec:app_models} for the details of PPISN Peak model; (iv). please remind that $f_{\rm dyn}$ and $r_{\rm 2G}$ (describing the mixing fractions) are also free parameters in the models; (v). we assume a redshift evolving merger rate of $R = R_0(1+z)^{2.7}$ and infer $R_0$ in the analysis.}
\end{table*}

\begin{table*}[htpb]
\begin{ruledtabular}
\caption{Hyperparameters, their descriptions, and choices of priors for the fiducial model in this Work}\label{tab:hp}
\begin{tabular}{ccc}
parameters & descriptions & priors \\
\hline
$\alpha$ & slope of the power-law underlying mass distribution of stellar BHs  & U(0,8) \\
$m_{\rm min}$ &minimum mass  of stellar-formed BHs &U(2,10)\\
$m_{\rm max}$ &maximum mass  of stellar-formed BHs &U(30,100)\\
$\delta_{\rm m}$& width of mass range that smoothing function impact on &U(0,10) \\
$\beta_{\rm field}$ & power-law index in the pairing function of field BBHs & U(0,6) \\
$\beta_{\rm dyn}$ & power-law index for mass ratio in the pairing function of dynamical BBHs & U(0,6) \\
$\gamma_{\rm dyn}$ & power-law index for total mass in the pairing function of dynamical BBHs & U(-4,12) \\
$f_{\rm dyn}$ & the mixing fraction of the dynamically formed BBHs &  U(0,1)\\
$\log_{10}r_{\rm 2G}$ & logarithm of the underlying fraction of the 2nd generation BHs &  U(-4,0)\\
\hline
$\mu_{\rm a,ste}$ & the Gaussian mean value for spin magnitude of stellar BHs & U(0,1) \\
$\sigma_{\rm a,ste}$ & the Gaussian width for spin magnitude of stellar BHs & U(0.05,0.5) \\
$\mu_{\rm a,2G}$ & the Gaussian mean value for spin magnitude of 2nd generation BHs & U(0,1) \\
$\sigma_{\rm a,2G}$ & the Gaussian width for spin magnitude of 2nd generation BHs & U(0.05,0.5) \\
$\sigma_{\rm ct,field}$ & the Gaussian width for spin orientation of the field BBHs & U(0.1,4) \\
$\sigma_{\rm ct,dyn}$ & the Gaussian width for spin orientation of the dynamical BBHs & U(0.1,4) \\
\hline
$\log_{10}r_0$ & logarithm of the local merger rate & U(0,3)\\
\end{tabular}
\tablenotetext{}{{\bf Note.} Here, `U' means uniform distribution.}
\end{ruledtabular}
\end{table*}

\section{Results}\label{sec:result}
We summarize in Tab.~\ref{tab:bf} the (logarithmic) Bayes factors of the models compared to the OnePL model. We interpret $\ln{\mathcal{B}} > 3.5$ ($\ln{\mathcal{B}} < -3.5$) as strong evidence for the corresponding model being more (less) supported by the data than the OnePL model. We also adopt the Akaike information criterion \cite[AIC=$2n-2\ln\mathcal{L}_{\rm max}$;][] {1981Likelihood} in the model comparison. The smaller the AIC, the greater the data’s preference for the model, and values of $\Delta \text{AIC}>6$ are considered significant for model selection \citep{2009ARNPS..59...95L}. The results show more support for the OnePL model, the TwoPL model, and the PPISN  model over the others, and their constraints on the hyper-parameters are shown in Appendix.~\ref{sec:app_post}.  

\begin{table*}[htpb]
\centering
\caption{Model comparison results}\label{tab:bf}
\begin{tabular}{lcc}
\hline
\hline
Models     &  $\ln{\mathcal{B}}$  &  $\Delta \text{AIC}$  \\
\hline
OnePL &0 &0 \\
TwoPL&-1.3 & 2.0\\
PPISN Peak &0.8& 2.8\\
OnePL \& IsoDyn  & -3.9 &11.0\\
OnePL \& Default  & -8.7 & 18.9\\
PP \& Default & -12.3 & 33.5\\
\hline
\hline
\end{tabular}
\end{table*}

\begin{figure*}
	\centering  
\includegraphics[width=0.48\linewidth]{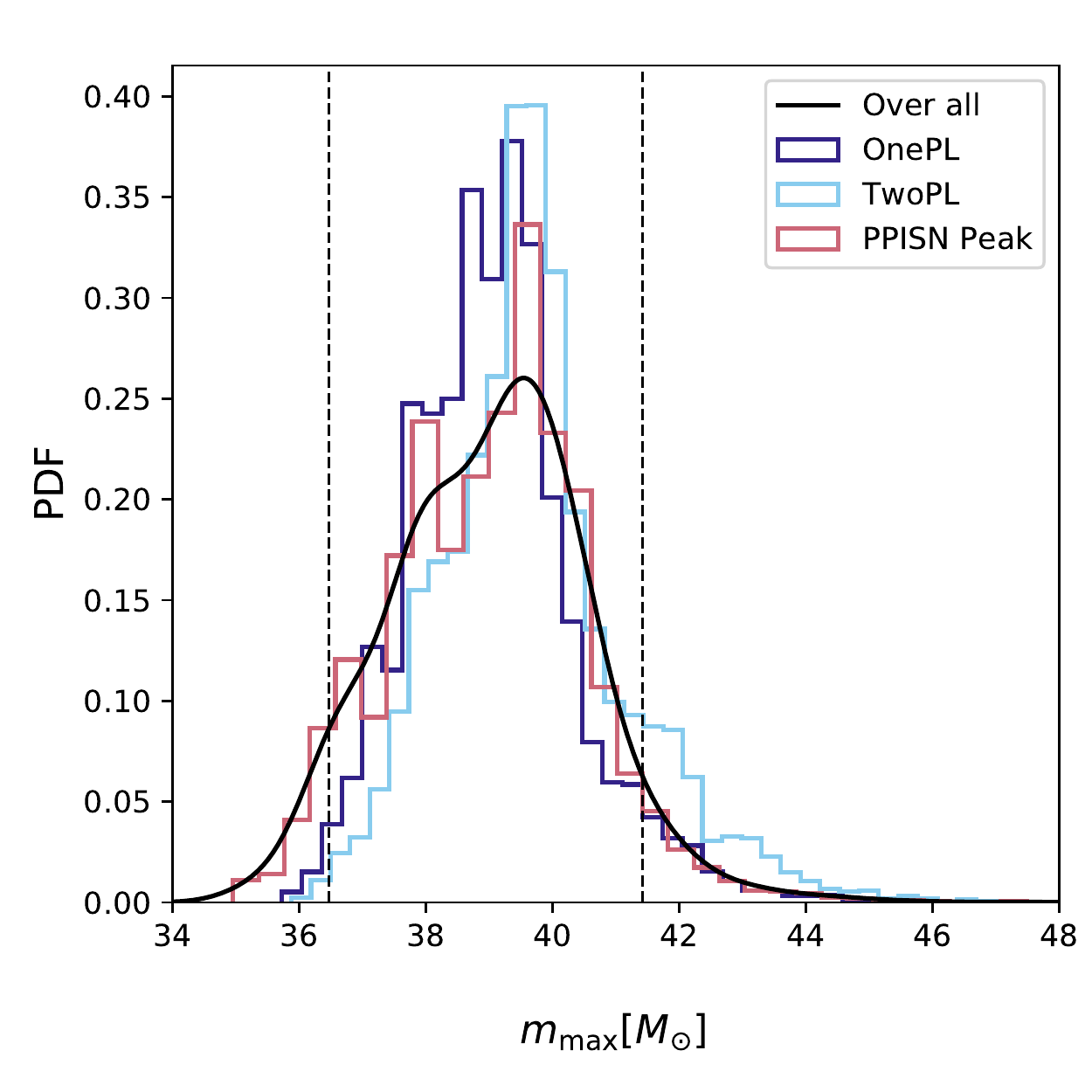}
\includegraphics[width=0.48\linewidth]{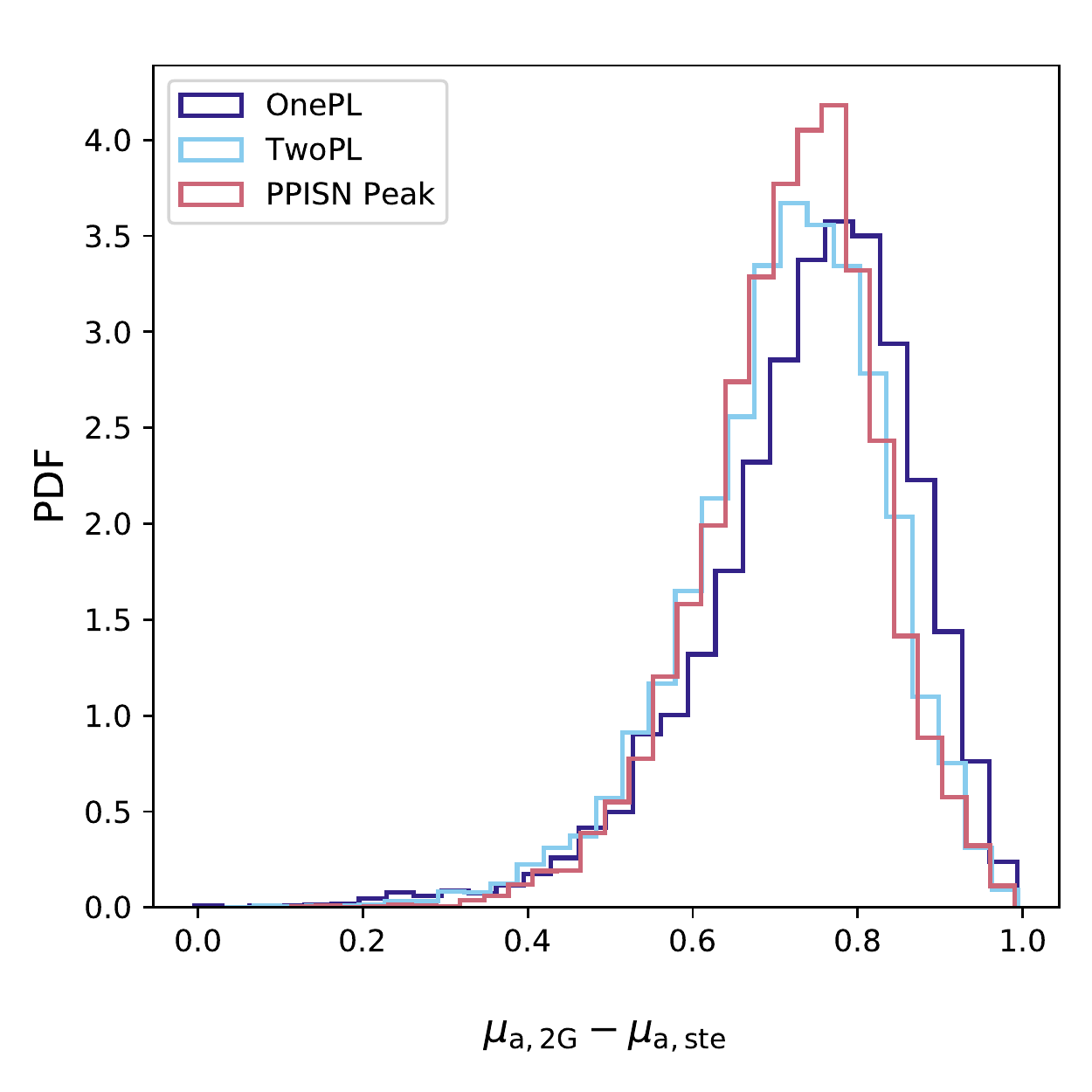}
\caption{The marginal posterior distributions for the maximum mass of field BBHs (left) and the difference between the mean values of spin magnitudes for the stellar-originated BHs and dynamical 2G BHs (right).}
\label{fig:mmax_dmua}
\end{figure*}

One intriguing result is that the $m_{\rm max}$ is tightly constrained in the inference. As demonstrated in Fig.~\ref{fig:mmax_dmua}, the marginal posterior distributions of $m_{\rm max}$ are consistent with each other across the three models. Combing the three posterior sample sets, we obtain an overall constraint of $m_{\rm max} = 39.1^{+2.4}_{-2.7}M_{\odot}$ at $90\%$ credibility. Such tight and consistent constraints may originate from the fact that $m_{\rm max}$ is also constrained by the secondary masses of BBHs, which are also forbidden from the high mass gap. We also investigate the possibility that $m_{\rm max}$ is mainly constrained by the measurement of the heaviest sample (GW190521), whose primary mass is about twice as large as $m_{\rm max}$. By leaving GW190521 out of the inference, we find that the change on the $m_{\rm max}$ posterior distribution is negligible. Thus, the constraints may come from the overall trend of data rather than a particular event. 

\begin{figure*}
	\centering  
\includegraphics[width=0.98\linewidth]{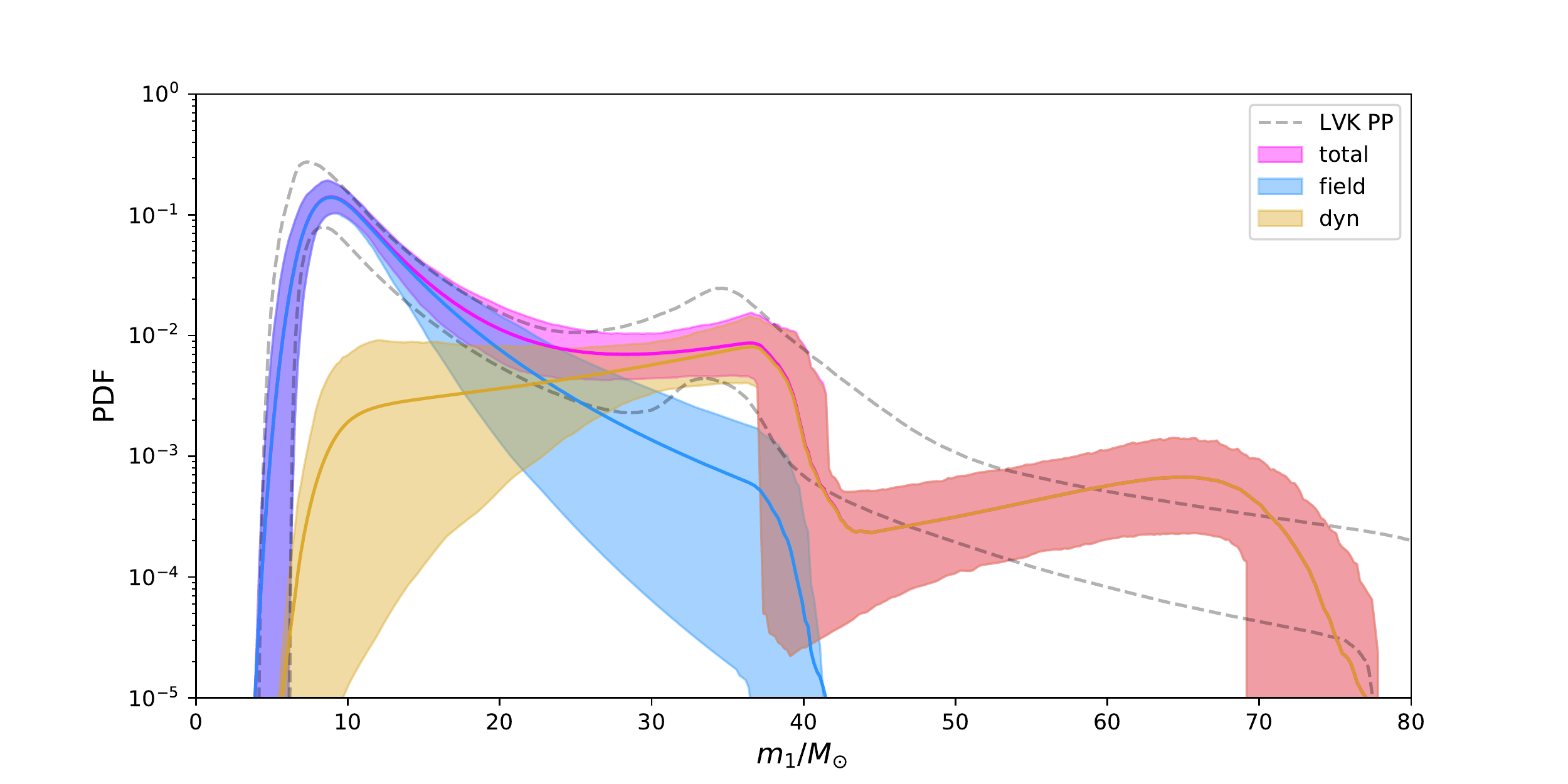}
\caption{The reconstructed astrophysical primary mass distribution for the OnePL model. The shaded areas indicate the $90\%$ credible regions for different components.}
\label{fig:m1}
\end{figure*}

\begin{figure*}
	\centering  
\includegraphics[width=0.98\linewidth]{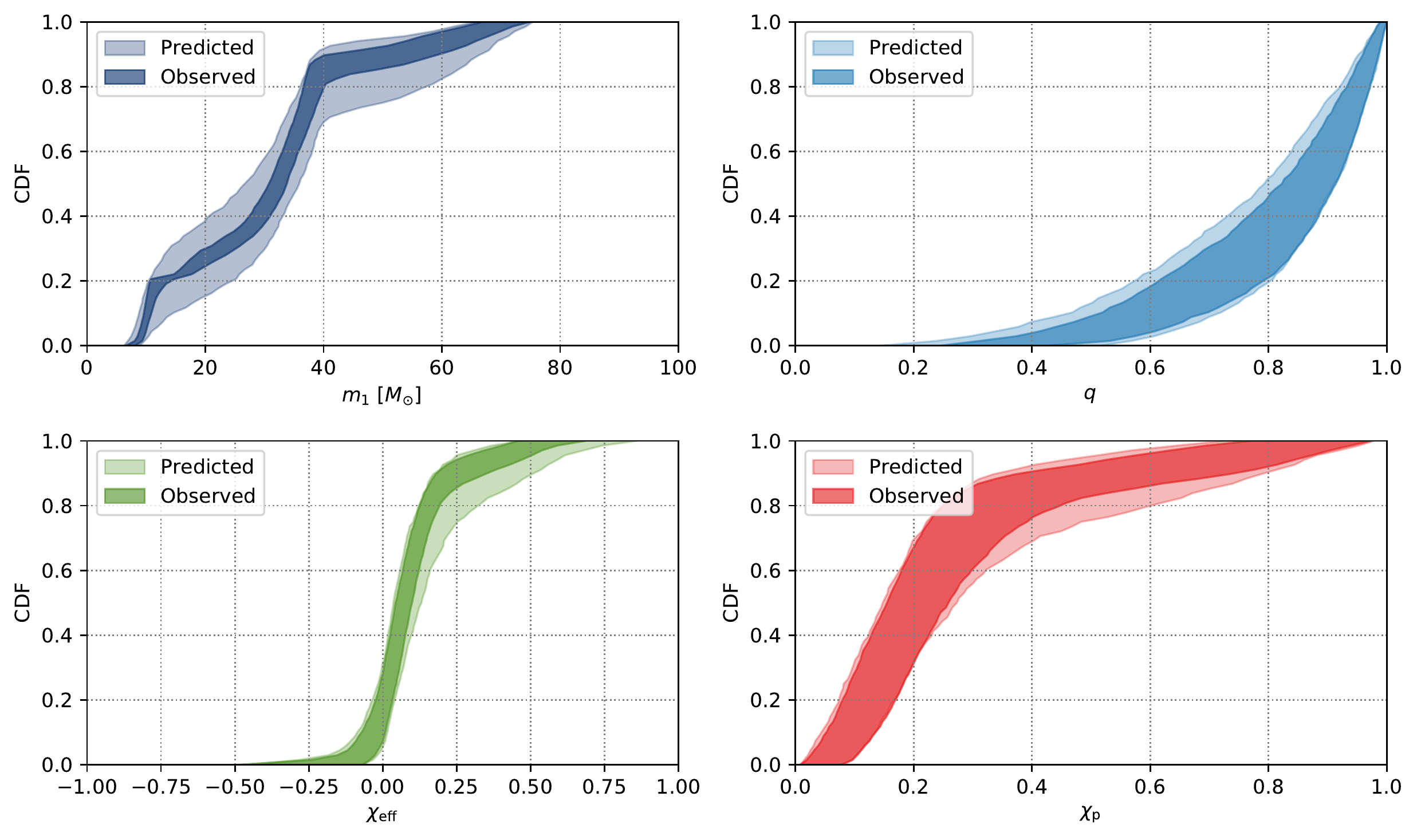}
\caption{Posterior predicted check for the reconstructed primary mass distribution (top-left), the mass ratio distribution (top-right), the effective inspiral spin distribution (bottom-left), and the effective precessing spin distribution (bottom-right).}
\label{fig:ppcheck}
\end{figure*}

To compare with the results in \cite{2021ApJ...913L...7A}, we plot the marginalized primary mass distribution in Fig.~\ref{fig:m1}. Unlike the $m_1$ distribution shown in Fig.~10 of \citet{2021arXiv211103634T}, the distribution inferred with our model shows a clearly sharp decay in $m_1 \sim 40 M_\odot$. The shallow bump extended to a higher mass is contributed by the 2G dynamical BHs. In some analyses on the GWTC-2 events, a similar sharp decay in the $m_1$ spectrum is also obtained using different approaches \citep{2021ApJ...913...42W,wong2021,2021ApJ...916L..16B}; now with GWTC-3, we confirm that our model with such a feature can better fit the data with respect to the PP \& Default model. To avoid model misspecification \citep{2022PASA...39...25R}, we perform posterior predicted checks following the procedures described in \citet{2021ApJ...913L...7A}. As shown in Fig.~\ref{fig:ppcheck}, the observed accumulated distribution of primary mass and mass ratio are within the predicted region of the model.

\begin{figure*}
	\centering  
\includegraphics[width=0.98\linewidth]{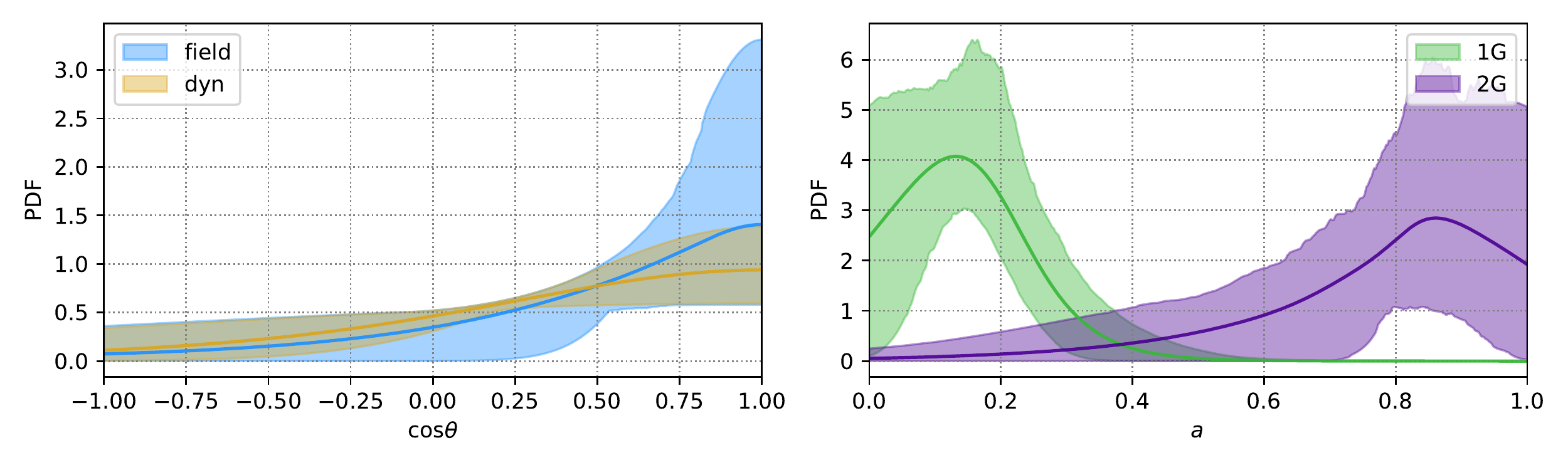}
\caption{The reconstructed astrophysical distributions for the cosine of the spin tilted angle (left) and the spin magnitude (right). The shaded areas indicate the $90\%$ credible regions.}
\label{fig:mua}
\end{figure*}

The inference also reveals a significant difference between the spin magnitude of the 2G dynamical BHs and stellar-origin BHs. For the OnePL model, we inferred that $\mu_{\rm a,ste} = 0.11^{+0.08}_{-0.09}$ and $\mu_{\rm a,2G} = 0.87^{+0.12}_{-0.24}$; these two parameters have consistent constraints in other models. We plot the difference between $\mu_{\rm a,ste}$ and $\mu_{\rm a,2G}$ in Fig.~\ref{fig:mmax_dmua}. Accounting for all models we have considered, the constraint on the difference is ${\rm d}\mu_{\rm a} > 0.4 $ at $99\%$ credibility. We note that for $\mu_{\rm a,ste}$, the posterior support does not go down to zero at $\mu_{\rm a,ste}=0$, while for $\mu_{\rm a,2G}$, having an average spin of $\sim 0.7$ (which is the case for the remnant of a coalescing system with non-spinning equal-mass components) is within its $90\%$ credible interval. We plot the $90\%$ credible region of the spin amplitude distribution in green for the field + 1G `dyn' mergers and in purple for the 2G `dyn' mergers in Fig.~\ref{fig:mua}. We also demonstrate in the figure the reconstructed distributions of $\cos{\theta_i}$ (which is governed by $\sigma_{\rm ct,i}$) of the field and dynamical mergers. The field channel shows more preference on the alignment of spin, although the evidence is not strong enough to claim the dynamical channel has a distinct $\cos{\theta_i}$ distribution due to the large uncertainties of the results. For both channels, the case of perfect alignment is not supported by the posterior distributions. The Bayes factor of the IsoDyn model compared to the OnePL model indicates that the isotropically tilted case for the dynamical BHs is also strongly disfavored. The posterior predicted check for the effective inspiral spin and the effective precessing spin are demonstrated in Fig.~\ref{fig:ppcheck}.

We address some caveats in interpreting the spin results: first, some astrophysical models predict more complex spin distributions (see Sec.~\ref{sec:astro}. for more discussions), and hence the simple truncated Gaussian has a limitation on recovering the exact shape of the distributions; however, our empirical model can still reflect the rough central tendency of data. Second, we have utilized parameter estimation samples derived with the default prior (Uniform distributions for $a_i$ and $\cos{\theta_i}$) for individual events. Thus our inference has limitations in identifying sub-populations with negligible spins, as demonstrated in \citet{2021ApJ...921L..15G}) (see however \citet{2022ApJ...937L..13C} and \citet{2022arXiv220902206T} for the arguments about this sub-population). We leave further investigation on this issue in future work since it will not change our current conclusion. 

The existence of a distinct underlying mass distribution for the dynamical BHs is ambiguous. The logarithmic Bayes factor of the TwoPL model compared to the OnePL model is $-1.3$, indicating the introduction of the additional parameter, $\alpha_{\rm dyn}$, does not significantly improve the model in describing the data. The posterior distribution of $\alpha_{\rm dyn}$ implies the underlying mass spectrum of this sub-population may be much flatter than that of the field BBHs, though their $90\%$ credible intervals overlap with each other in the range of $3.17-4.16$.

\begin{figure*}
	\centering  
\includegraphics[width=0.8\linewidth]{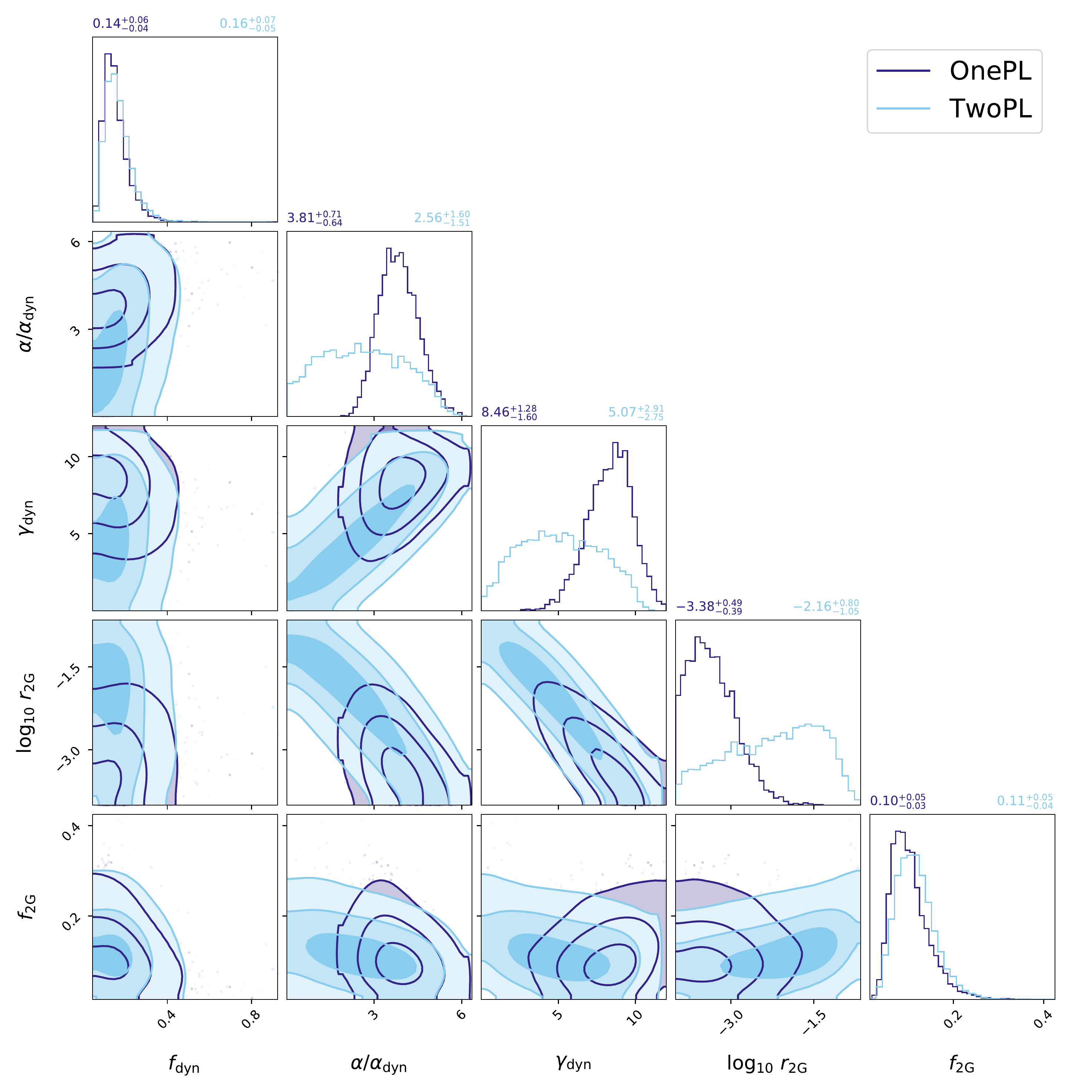}
\caption{Posterior distribution the parameters about the dynamical channel for the OnePL and the TwoPL models; note that $r_{\rm 2G}$ is the fraction of 2G BHs in the dynamical environment, and $f_{\rm 2G}$ is the fraction of BBHs involving a least one 2G BH.}
\label{fig:correlate}
\end{figure*}

The results also reveal different pairing functions between the `field' BHs and `dyn' BHs. For the OnePL model, we infer $\gamma_{\rm dyn} = 8.41^{+2.15}_{-2.60}$; for the TwoPL model, the constraint is much weaker, and $\gamma_{\rm dyn}$ strongly degenerates with $\alpha_{\rm dyn}$ and $\log_{10} r_{\rm 2G}$ in the inference. We show the corner plot for these parameters in Fig.~\ref{fig:correlate} to illustrate the degeneracy. The parameter $\beta_{\rm dyn}$ is constrained to be $3.56^{+2.41}_{-1.71}$ ($3.54^{+3.39}_{-1.89}$) for the OnePL (TwoPL) model; for the pairing function of field binaries, the posterior distributions of $\beta_{\rm field}$  skew toward small values. This finding is consistent with the results obtained by \citet{2022ApJ...933L..14L} that suggests the events with larger masses (likely dominated by dynamical mergers) prefer more symmetric systems. As a conclusion for the pairing function, one requirement is clear for our model to fit the data: the underlying dynamical BHs tend to pair together with more equal masses and larger total mass.

We further use the results to derive branch ratios for different types of mergers. Combing the posterior distributions from OnePL, TwoPL, and PPISN models, we obtain $f_{\rm dyn} = 0.15^{+0.12}_{-0.6}$, which means $\sim 15\%$ of the astrophysical mergers would form through the dynamical process. The parameter $\log_{10} r_{\rm 2G}$ describes the fraction of underlying 2G BHs in the dynamical environments. Its posterior distribution for the TwoPL model deviates from the results of the OnePL model and the PPISN model due to its degeneracy with $\gamma_{\rm M}$ and $\alpha_{\rm dyn}$. We further translate the constraint on $r_{\rm 2G}$ into the fraction of mergers that contain at least one 2G BH in the dynamical sup-population, and the result is shown in Fig.~\ref{fig:correlate}. Combing the constraints from different models, the fraction is $0.10^{+0.06}_{-0.03}$ at $90\%$ credible level. After accounting for the selection effect in the observation, our result suggests that such events make up $16^{+8}_{-7}\%$ of the detectable mergers, and $6^{+6}_{-4}\%$ of the systems have both components being 2G BHs. We also compute the probability of the formation channels for each event according to the constrained OnePL model (details are presented in Appendix.~\ref{sec:channels}), and find that 11 out of the 69 events are most likely (probability $>50\%$) containing 2G BHs. In particular, GW190521\_030229 and GW191109\_010717 have  probabilities of $\sim 1$ and 0.74 to be 2G+2G BBHs.

\section{The Spin-dominant Analysis}\label{sec:spin}
We have shown in Sec.\ref{sec:result} that the current data can be explained by our model with two sub-populations. Two important constraints on our model were derived: the $m_{\rm max} \sim 40 M_\odot$ and the significant difference between the spin magnitudes of the two sub-populations. To investigate the model dependency of these results, here we perform the ``spin-dominant analysis". The model used in the analysis has consistent astrophysical motivation with the more complex models in Sec.\ref{sec:model}, while its parameters are free from the assumptions of the mass model.

We assume that the BHs with masses above a certain division ($m_{\rm d}$) have a different spin distribution from that of BHs below $m_{\rm d}$, then our model describing an individual BH can be written as:
\begin{equation}\label{eq:app_spin}
\pi_{\rm s}(a_i,\cos{\theta_i},m_i|\mathbf{\Lambda}_{\rm s}) = 
\begin{cases}
\mathcal{G}(a_i|\mu_{\rm a,1},\sigma_{\rm a,1})\mathcal{G}'(\cos{\theta_i}|1,\sigma_{\rm ct,1}) & ~\text{for}~m_i<m_{\rm d}\\
\mathcal{G}(a_i|\mu_{\rm a,2},\sigma_{\rm a,2})\mathcal{G}'(\cos{\theta_i}|1,\sigma_{\rm ct,2}) & ~\text{for}~m_i>m_{\rm d}\\
\end{cases}
\end{equation}
the parameters included in $\gamma_{\rm s}$ are summarized in Tab.~\ref{app:spin_prior} along with their descriptions and priors. Note that both BHs in a system follow this model, so the likelihood for a particular posterior sample of an event is $\pi_{\rm s}(a_1,\cos{\theta_1},m_1|\mathbf{\Lambda}_{\rm s}) \times \pi_{\rm s}(a_2,\cos{\theta_2},m_2|\mathbf{\Lambda}_{\rm s})$. Though different sub-populations may have overlapping BH masses, the model can represent a switch of the dominant sub-population at $m_{\rm d}$.

\begin{table}[htpb]\label{app:spin_prior}
\begin{ruledtabular}
\caption{Hyperparameters, their descriptions, and priors for the spin model described by Eq.~(\ref{eq:app_spin})}
\begin{tabular}{ccc}
parameters & descriptions & priors \\
\cline{1-3}
$m_{\rm d}[M_{\odot}]$ & the division point in the mass range that diving the two subpopulations  & U(20,70) \\
$\mu_{\rm a,1}$ & the Gaussian mean value for the spin magnitude of the BHs below $m_{\rm d}$& U(0,1) \\
$\sigma_{\rm a,1}$ & the Gaussian width for the spin magnitude of stellar the BHs below $m_{\rm d}$ & U(0.05,0.5) \\
$\sigma_{\rm ct,1}$ & the Gaussian width for spin orientation of the BHs below $m_{\rm d}$ & U(0.1,4) \\
$\mu_{\rm a,2}$ & the Gaussian mean value for the spin magnitude of the BHs above $m_{\rm d}$ & U(0,1) \\
$\sigma_{\rm a,2}$ & the Gaussian width for spin magnitude of the BHs above $m_{\rm d}$ & U(0.05,0.5) \\
$\sigma_{\rm ct,2}$ & the Gaussian width for spin orientation of the BHs above $m_{\rm d}$ & U(0.1,4) \\
\end{tabular}
\tablenotetext{}{{\bf Note.} Here, `U' means uniform distribution.}
\end{ruledtabular}
\end{table}

\begin{figure}
	\centering  
\includegraphics[width=0.98\linewidth]{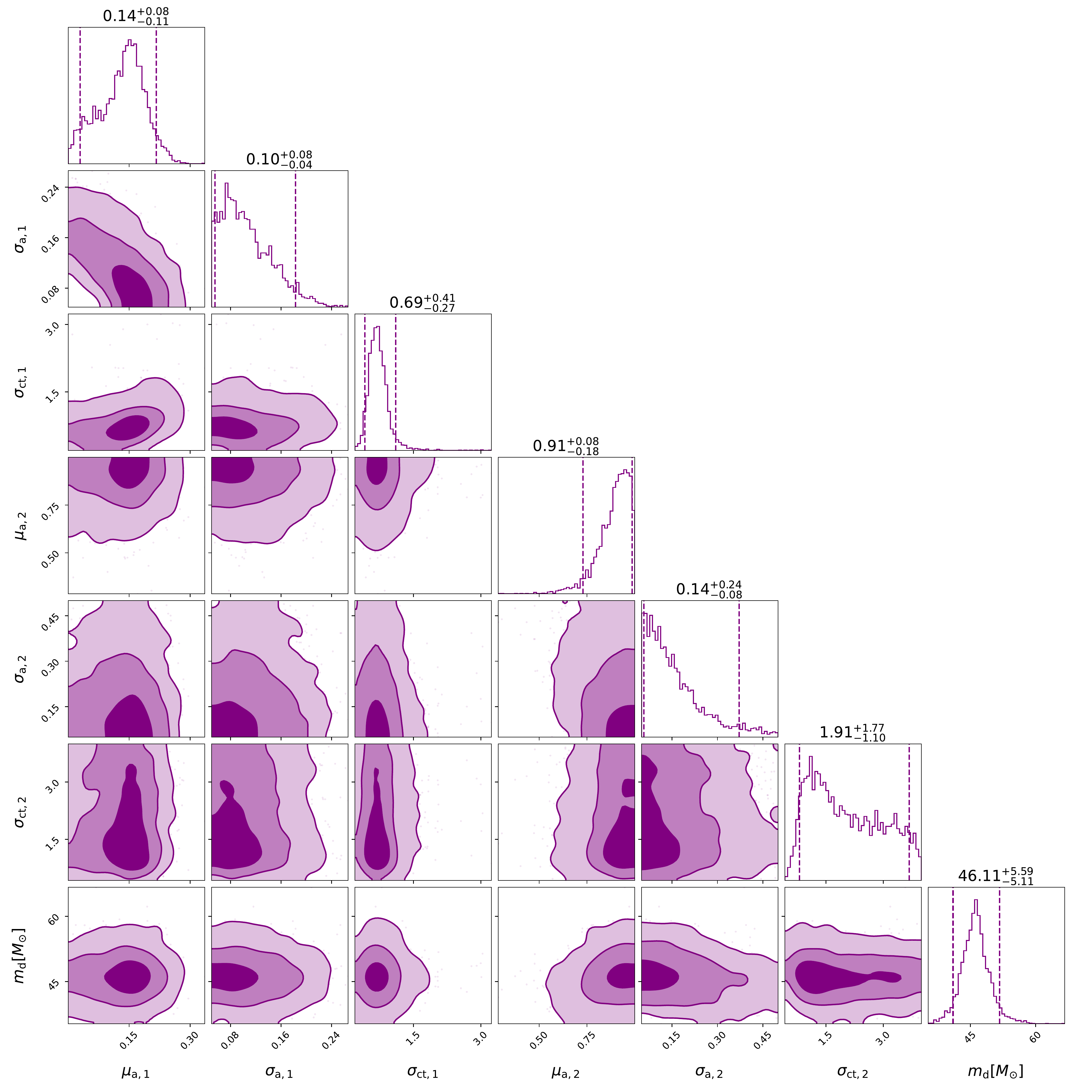}
\caption{Posterior distributions of parameters of the spin model described by Eq.(\ref{eq:app_spin}), the dashed lines represent the 90\% credible intervals.}
\label{fig:app_spin}
\end{figure}

The inferred hyper-parameters for this model are shown in Fig.~\ref{fig:app_spin}. We find that the constraints for ($\mu_{\rm a,1}$, $\sigma_{\rm a,1}$, $\sigma_{\rm ct,1}$, $\mu_{\rm a,2}$, $\sigma_{\rm a,2}$, $\sigma_{\rm ct,2}$ ) are consistent with those for ($\mu_{\rm a,ste}$, $\sigma_{\rm a,ste}$, $\sigma_{\rm ct,field}$, $\mu_{\rm a,2G}$, $\sigma_{\rm a,2G}$, $\sigma_{\rm ct,dyn}$) in the OnePL model. We derive a division mass of $m_{\rm d} = 46.11^{+5.59}_{-5.11}M_{\odot}$, which is in agreement with the inferred $m_{\rm max}$ but with a slightly larger median. It is worth noting that neglecting the spin data has a very small impact on the constraints on $m_{\rm max}$, so we conclude that both mass and spin data show evidence that the dominant sub-population may change at a mass of $\sim 36-52 M_\odot$.

\section{Astrophysical Implication}\label{sec:astro}

Our phenomenological models provide a possible way to fit the observed BBH population. Here we discuss the question: what astrophysical conditions are required to produce the resulting sub-populations (described by the constrained hyper-parameters)? This is a self-consistency check for our model; alternatively, assuming our model is correct, this might provide insights into some key aspects regarding the formation of BBHs.

In principle, the position of the maximum BH mass can be explained by a combination of PI physics at low metallicities (Z) and Fe-dependent Wolf-Rayet (WR) mass loss at high Z without needing to change the uncertain nuclear reaction rate. As shown in \citet{2019ApJ...887...53F,farmer2020}, a $M_{\rm max} \sim 40 M_\odot$ can be attained by varying the $^{12}\mathrm{C}(\alpha, \gamma)^{16}\mathrm{O}$ thermonuclear reaction rate within its $1 \sigma$ uncertainties for the progenitors with $Z \leq 3\times 10^{-3}$. On the other hand, at higher $Z$, iron (Fe) dependent WR winds \citep{Vink:2005zf,Sander:2020amn} could be the leading factor that decides $M_{\rm max}$. When the recent WR theoretical mass-loss recipe of \citet{Sander:2020amn} was employed in stellar models by \citet{Higgins:2021jux}, it was found (Fig.~13 in that paper) that above $Z/Z_{\odot} = 0.5$ the maximum final mass was of order $40\,M_{\odot}$, and close to the maximum BH mass uncovered by our phenomenological model.

When a more complex function of the (maximum) BH masses becomes available as a function of $Z$ and redshift $z$, the issue of constraining the $^{12}\mathrm{C}(\alpha, \gamma)^{16}\mathrm{O}$ thermonuclear reaction rate could be revisited. In addition, the sub-population of very hot WR stars and the WO stars could help constrain this elusive nuclear reaction rate \citep{2015A&A...581A.110T}. In other words, multi-messenger astrophysics, including both GW events and electromagnetic (EM) information, could be combined to determine both the $^{12}\mathrm{C}(\alpha, \gamma)^{16}\mathrm{O}$ rate as well as stellar wind physics, by disentangling high $Z$ and low $Z$ GW events.

The BH spins have been widely considered to distinguish the BBH formation channels. In the isolated evolution of massive stars via common-envelope phase \citep{2013A&ARv..21...59I}, the first-born BH originating from an initially massive star has negligible/small spin due to the standard angular momentum transport \citep{2018A&A...616A..28Q,2019MNRAS.485.3661F,2020A&A...636A.104B}; the second-born BH from the less massive star can have spin from zero to maximally spinning due to tidal spin-up \citep{2018A&A...616A..28Q,2022ApJ...928..163H}, while this effect is limited within the horizon of current gravitational-wave detectors \citep{bavera2020,2022arXiv220402619B}. These predictions seem to be in line with our inferred result which allows for the `field' sub-population to have $\mu_{\rm a,ste} \sim \sigma_{\rm a,ste} \sim 0.1$. Nevertheless, within their $90\%$ credible region, $\mu_{\rm a,ste}$ and $\sigma_{\rm a,ste}$ have more posterior supports at slightly larger values. Higher spins in isolated binary evolution can be attained by various processes, like the tidal spin-up of two He stars following the double-core common envelope phases \citep{2021ApJ...921L...2O}, the Eddington-limited accretion onto BHs, the efficient angular momentum transport within massive stars \citep{2022ApJ...924..129Q,2022arXiv220302515Z}, and the chemically homogeneous evolution \citep{demink2016,2016A&A...588A..50M}. In addition, both the OnePL/TwoPL/PPISN Peak model and the spin-dominant analysis show that the spin-tilted angle of ``field" BBHs are not perfectly aligned.  The spins may be tossed due to the supernova kicks during their formation process in the core collapse of massive stars \citep{2022arXiv220502541T}.

We have inferred a total local ($z=0$) BBH merger rate of $18.2^{+7.65}_{-5.4}~{\rm Gpc^{-1} yr^{-1}}$, consisting with the result in \citet{2021arXiv211103634T}. Our models enable us to derive branch ratios for different sub-populations. The analysis has shown that the dynamical channel contributes to $9\% - 27\%$ of the mergers, in agreement with the expectations of some dynamical environments \citep{2021NatAs...5..749G}. The investigation of spin distributions in this work sheds light on the details of dynamical environments. Dynamical assembly in dense stellar environments like globular clusters predict isotropic spins, while the features for the mergers in the AGN disks can change in a variety of ranges depending on the properties of the disk \citep{2022ApJ...931...82V} and the dynamics in play \citep{2020ApJ...899...26T}. Though at this stage, we might not be able to clearly resolve the tilted angle distribution from the available data \citep{2022arXiv220902206T,2022arXiv220906978V}, our analysis disfavors both the perfectly alignment case ($\sigma_{\rm ct,dyn} \ll 1$) and the isotropically tilted case ($\sigma_{\rm ct,dyn} \gg 1$) of the spin, implying that not all of the dynamical mergers happen in old, dense AGNs, or, young, dilute AGNs/globular clusters \citep{2013LRR....16....4B,2022ApJ...931...82V}. It is possible that multiple dynamical environments are contributing to the observable events, and our results suggest that some of the events are likely from AGN disks. The reconstructed spin amplitude distribution for the 2G BBHs has $\mu_{\rm a,2G} \gg 0$, which can be explained by the inheritance of orbital angular momentum of 1G mergers \citep{fishbach2017} in the framework of our model. For a coalescing system with non-spinning equal-mass components, the remnant will have spin amplitude $\sim 0.7$; On the one hand, non-zero and isotropically distributed spins of 1G BHs (due to, e.g., natal kicks) will increase the dispersion of spin amplitude distribution of 2G mergers; on the other hand, if the component spins and orbital angular momentum were aligned by the accretion torque and co-rotation torques in the AGN disk, the remnants could have larger spins. Recently, \citet{2022ApJ...935L..26F} addressed the limits on hierarchical black hole mergers from the most negative $\chi_{\rm eff}$ systems; their work mainly focus on classical star clusters, further studies can be carried out to quantitatively constraint the mergers in an AGN disk.

Clues for the Hierarchical sub-populations was suggested by \citet{2021ApJ...915L..35K} using the GWTC-2 data. With the GWTC-3 events and by employing the pairing function in the model, we have inferred sub-populations with distributions different from their work. The pairing function for the dynamical sub-population prefers systems with heavier total mass, though large systematical differences are presented in the posterior distributions of $\gamma_{\rm dyn}$ between the OnePL and the TwoPL model due to its degeneracy with $\alpha_{\rm dyn}$ in the inference. \citet{oleary2016} found that dynamical interactions can enhance the merger rate of BBHs by a boost factor $\propto M_{\rm tot}^{\gamma}$, with $\gamma \geq 4$, which is consistent with the constraints on $\gamma_{\rm dyn}$ in our model. It is interesting to point out \citet{yang2019} addressed the AGN disk will also harden the mass distribution of BBHs, and the power-law index of the mass spectrum will change by $\Delta \alpha_{\rm dyn} \sim 1.3$. This possibility is included in the results for the TwoPL model, considering the statistical uncertainties. The constraint on $\beta_{\rm dyn}$ is in agreement with the conclusions from some N-body simulations that the pairing of BHs of comparable masses is energetically favorable \citep{rodriguez2016,banerjee2017}. Further studies comparing the theoretical predictions with the constraints on $\gamma_{\rm dyn}$ and $\beta_{\rm dyn}$ could help better understand the dynamic interactions in dense environments.

Among the 11 events (shown in bold in Tab.~\ref{tab:pchannel}) plausibly contain 2G BHs, the possible hierarchical origins for GW190519\_153544, GW190521\_030229, GW190602\_175927 and GW190706\_222641 have also been suggested in \citet{2021ApJ...915L..35K}. GW191109\_010717, which has a probability of 0.74 to contain two 2G BHs, has non-negligible orbital eccentricity and precession \citep{2022arXiv220614695R,2022CQGra..39l5003H}.
Additionally, evidence for the precession in GW190929\_012149 was reported by \citet{2022PhRvD.106b3019H}.
Nevertheless, \citet{vigna2021} proposed that the progenitor of GW170729\_185629 is a low-metallicity field triple. Further studies on the signals or properties of these individual events may provide additional clues for their origins.

\section{Summary and Discussion}\label{sec:summary}
In this work, we study the joint mass-spin distribution of coalescing binary black holes by incorporating the pairing functions into the hierarchical Bayesian inference. With astrophysical concerns, we propose a model consisting of two sub-populations having different pairing functions and representing binaries form in the filed and dynamical environments, respectively. We find that this model can well explain the current population of BBHs observed by LIGO/Virgo/KAGRA. Under the framework of our model, some key features of different formation channels are revealed and well constrained in the analysis. For the field binaries and 1G dynamical binaries, we obtain a tight constraint on the maximum mass of $m_{\rm max} = 39.1^{+2.4}_{-2.7} M_\odot$; we also infer that the 2G dynamical mergers have an average spin magnitude that is significantly larger than other kinds of mergers, with ${\rm d}\mu_{\rm a} > 0.4 $ at $99\%$ credibility. As we have discussed in Sec.~\ref{sec:astro}, the inferred values of hyper-parameters in our phenomenological model can be naturally explained by current astrophysical theories. Revealing the potential features and assembling tendency in the mass and spin distributions can help distinguish the acting mechanisms during the formation of field BBHs, as well as the environments of dynamical BBHs.

Our model in this work is designed to capture the main features in the distributions of field binaries and dynamical binaries if they exist; it could be further improved and extended in the future to identify sub-classes under each sub-population. Very massive BHs lying in the conventional PI gap can also be formed through the preserve of large H envelope for progenitor stars with reduced metallicity ($Z< 0.1 ~ Z_{\odot}$ or below) \citep{2021MNRAS.504..146V}, and could be responsible for some events like GW\,190521. Field binaries may also yield systems with relatively high spins and large masses (below the mass gap) through chemically homogeneous evolution \citep{mandel2016,2016A&A...588A..50M,dubuisson2020,2022A&A...657L...8B,2022arXiv221105945Q}. Further studies with additional considerations on these channels are needed to clarify their existence and branch ratios in the GW events. 

With more events from future observations, the redshift evolution of mass function and spin distributions might be well modeled, and there are already some studies \citep{2021ApJ...912...98F,2022ApJ...932L..19B} attempting to reveal the trends with current data. Features in the mass spectrum of compact binaries via GW observations, especially the high-mass gap \citep{2020ApJ...904L..26F} and the low-mass gap \citep{2021ApJ...923...97L,2022ApJ...937...73Y}, are great ingredients in the ``spectrum siren" for cosmological probe \citep{2022PhRvL.129f1102E}. Therefore, extending our method to study the redshift evolution of the maximum mass for the BHs may bring new insight into the constraints on the cosmic expansion history \citep{2019ApJ...883L..42F}.

\acknowledgments
 This work was supported in part by the National Natural Science Foundation of China (NSFC) under grants No. 11921003, No. 12203101, No. 11773078, and No. 11933010. 
 YQ acknowledges the support from NSFC (Grant Nos. 12003002, 12192220, 12192221) and the Natural Science Foundation of Universities in Anhui Province (Grant No. KJ2021A0106).

\vspace{5mm}
\software{Bilby \citep[version 1.1.4, ascl:1901.011, \url{https://git.ligo.org/lscsoft/bilby/}]{2019ascl.soft01011A},
          PyMultiNest \citep[version 2.11, ascl:1606.005, \url{https://github.com/JohannesBuchner/PyMultiNest}]{2016ascl.soft06005B}.
          }

\bibliographystyle{aasjournal}
\bibliography{bibliography}

\appendix
\section{modifications of the population model}\label{sec:app_models}
\begin{figure}
	\centering  
\includegraphics[width=0.98\linewidth]{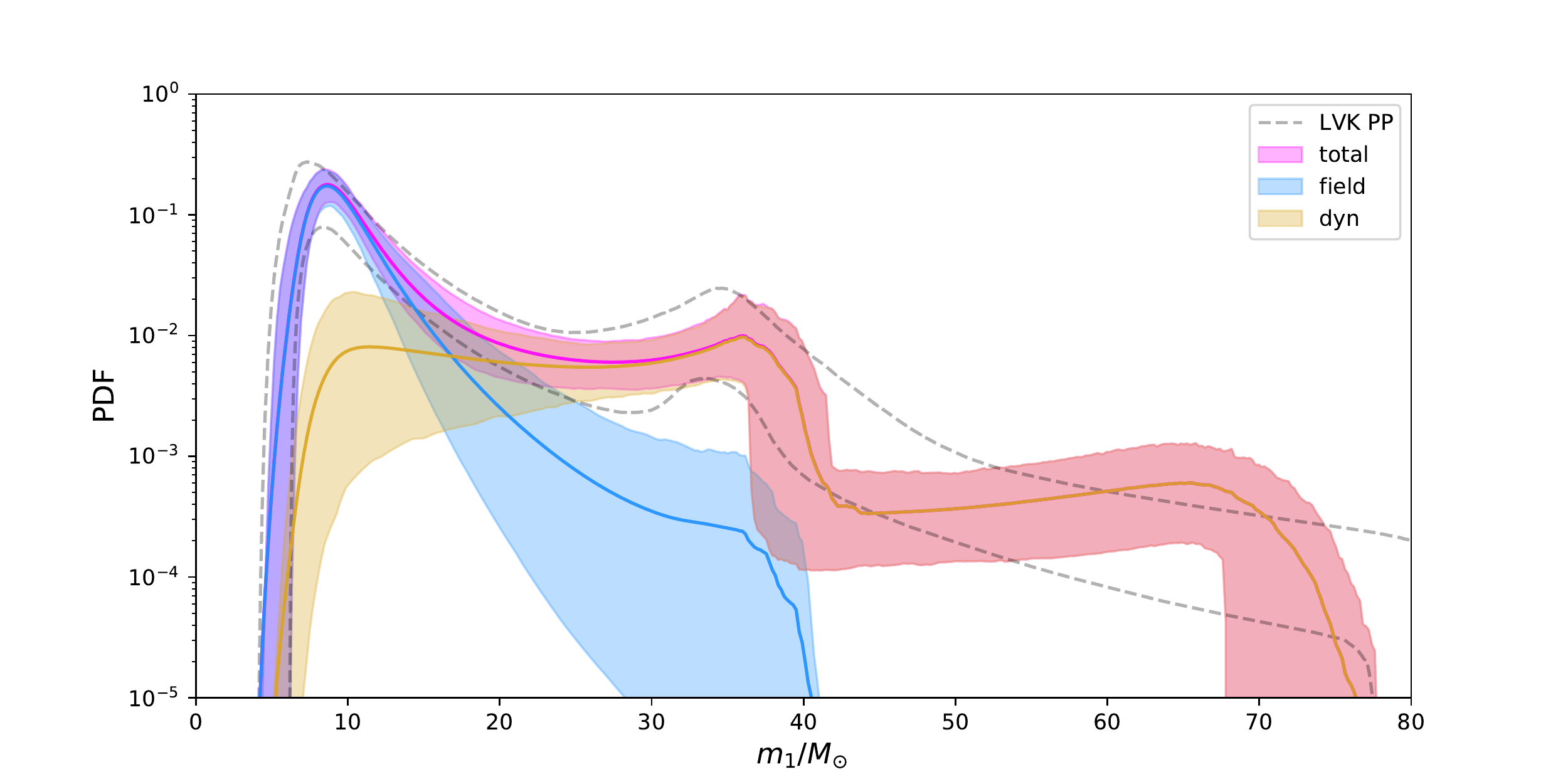}
\caption{The same as Fig.\ref{fig:m1}, but for the PPISN Peak model.}
\label{fig:ppsin_m1}
\end{figure}

We also try to take some widely concerned astrophysical motivations into our population models. Firstly, we consider the pile-up at the high-mass cutoff of the BH mass spectrum that results from the PPISNe (labeled 'PPISN Peak') \citep{2018ApJ...856..173T},
\begin{equation}\label{eq:app1}
\begin{aligned}
\pi_{\rm ste,m}(m_i|\mathbf{\Lambda}_{\rm ste}) &= \mathcal{P}(m_i|\alpha, m_{\rm min}, m_{\rm max}, \delta_{\rm m})*(1-r)+r*\mathcal{G}(m_i|\mu,\sigma,m_{\rm min},m_{\rm max}),\\
r&=\int_{m_{\rm max}}^{m_{\rm max}+\Delta_{\rm m}}{\text{d}m\mathcal{P}(m|\alpha, m_{\rm min}, m_{\rm max}+\Delta_{\rm m}, \delta_{\rm m})}
\end{aligned}
\end{equation}
where $\mathcal{P}$ is the normalized power law with smoothing treatment on the low-mass cutoff, so $r$ represents the fraction of the underlying BHs that could have been in the range of $\Delta_{\rm m}$ above $m_{\rm max}$, but appeared below $m_{\rm max}$ because of the PPISN. 
Note that $\Delta_{\rm m}$ should be smaller than the width of the high-mass gap caused by the (P)PISN. Therefore, we set the prior of $\Delta_{\rm m}$ to be uniformly distributed in $(0,100)M_{\odot}$, $\Delta_{\rm m}=100M_{\odot}$, which is wider than the widely expected mass gap \citep{2017MNRAS.470.4739S}.
As presented in Fig. \ref{fig:ppsin_m1}, the contribution of the PPISN channel is negligible, and the high-mass cutoff is also tightly constrained, as shown in Fig.\ref{fig:mmax_dmua}. However, the $\Delta_{\rm m}$ is not well constrained (i.e., is similar to the prior distribution, and $\Delta_{\rm m}=0M_{\odot}$ ($\Delta_{\rm m}=100M_{\odot}$) can not be ruled out).

The BHs formed through different formation channels/environments may follow different spectrum indexes.
Therefore, we introduce the second modified model (labeled `TwoPL'), in which we consider another power-law index $\alpha_{\rm dyn}$ for the mass function of first-generation BHs in the dynamical assembly. 

\begin{table}[htpb]
\begin{ruledtabular}
\caption{The additional hyper-parameters, and priors for the modified model}\label{tab:app_prior}
\begin{tabular}{cccc}
Models   & parameters & descriptions & priors \\
\cline{1-4}
\multirow{3}{*}{PPISN Peak} & $\Delta_{\rm m}$ & the mass range above $m_{\rm max}$ that contribute to the PPISN peak & U(0,100) \\
& $\mu$ & the Gaussian central value of the PPISN peak & U(20,50) \\
& $\sigma$ & the Gaussian width of the PPISN peak & U(1,10) \\
\end{tabular}
\tablenotetext{}{{\bf Note.} Here, `U' means the uniform distribution.}
\end{ruledtabular}
\end{table}

Finally, we consider the spin orientation of BBHs formed by the dynamical capture may have isotropic distribution; therefore, in the modified model (labeled ``IsoDyn"), we assume that $\pi_{\rm dyn}(\cos{\theta_{1,2}})=1/2~\text{for}~\cos{\theta_{1,2}}\in(-1,1)$, i.e., $\cos{\theta_{1,2}}$ uniformly distributed in (-1,1). However, from the results summarized in Tab.\ref{tab:bf}, such a scenario is not favored by the GW observation.

\section{Full corner plot for The posterior distributions}\label{sec:app_post}

\begin{figure*}
	\centering  
\includegraphics[width=1.\linewidth]{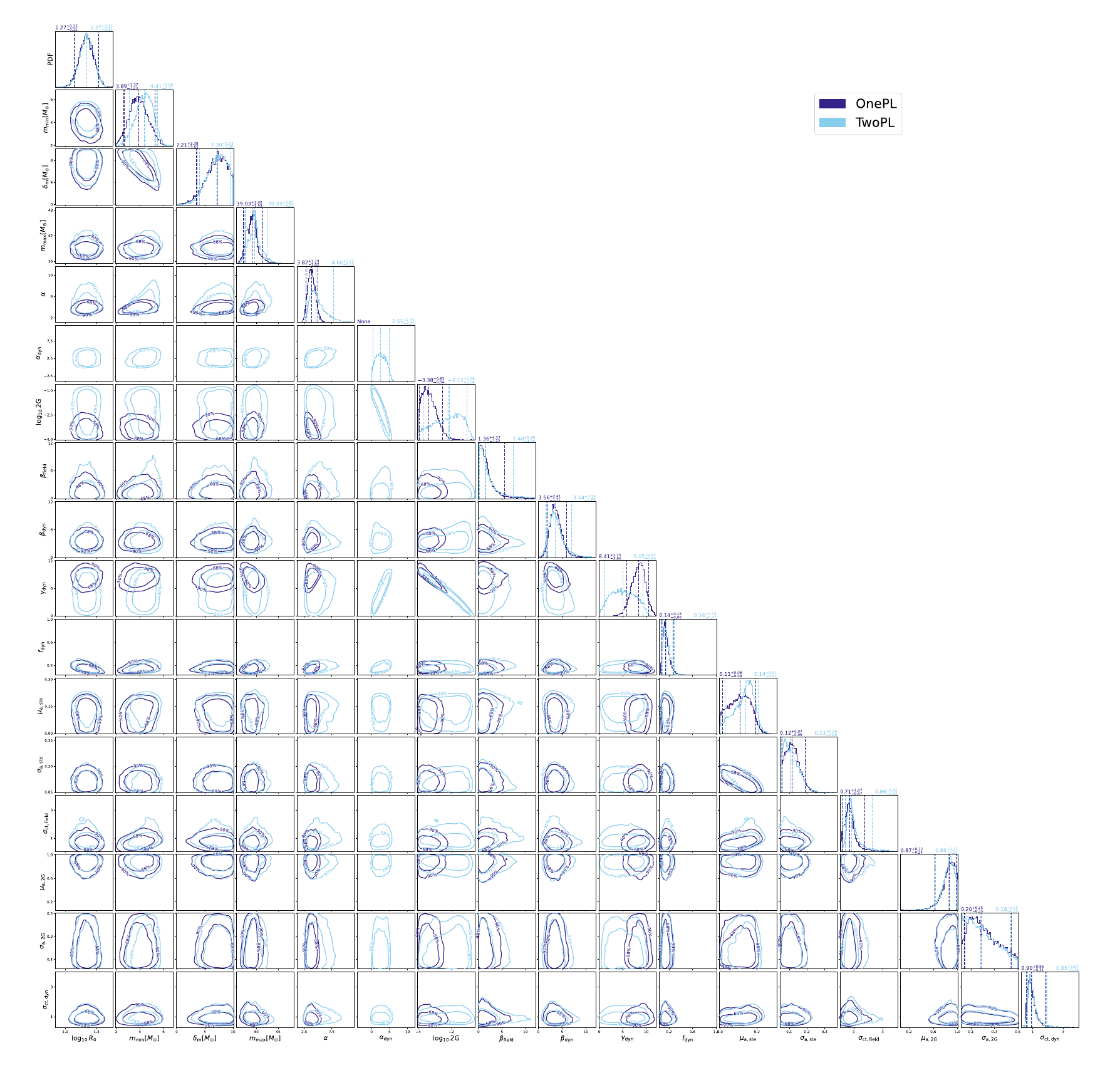}
\caption{Posterior distribution of all the parameters obtained by the OnePL and the TwoPL models; the dashed lines in the marginal distribution represent the $90\%$ credible intervals. }
\label{fig:full}
\end{figure*}

We compare the inferred parameters obtained by the OnePL model and the TwoPL model in Fig.~\ref{fig:full}. The PPISN Peak model has similar constraints for the overlapping parameters with the OnePL model.  We find that most parameters are nearly identical in both models except for $\gamma_{\rm dyn}$, $\log_{10}r_{\rm 2G}$, which are strongly degenerate with $\alpha_{\rm dyn}$. Though the constraint on $\log_{10}r_{\rm 2G}$ is weak, as shown in Fig.~\ref{fig:correlate}, the fraction of the mergers that involve at least one 2nd-generation BH is well constrained.

\section{Probabilities of formation channels for each event}\label{sec:channels}

Using the constraints on the OnePL model, we calculate the probabilities for the formation channels (field, dynamical 1G+1G, 1G+2G\footnote{Here 1G+2G means a 1G primary BH and a 2G secondary BH.}, 2G+1G, and 2G+2G) of each event. With the inferred posterior distribution $p({\bf \Lambda}|\vec{d})$, the probability of the $i$-th event came from $K$ channel is
\begin{equation}
P_{K}^{i}\equiv \frac{\int{{\rm d}{\bf \Lambda}Z(d_i|{\bf \Lambda}, K)r_{K}({\bf \Lambda})p({\bf \Lambda}|\vec{d})}}{\int{{\rm d}{\bf \Lambda}Z(d_i|{\bf \Lambda})p({\bf \Lambda}|\vec{d})}}
\end{equation}
where $r_K$ is the mixing fraction of the $K$ channel, $Z(d_i|{\bf \Lambda}, K)$ is the evidence for $K$ channel, and $Z(d_i|{\bf \Lambda})$ is the evidence for the combination of all channels, which read
\begin{equation}
Z(d_i|{\bf \Lambda}, K)=\int{{\rm d}{\bf \theta}\mathcal{L}(d_i|{\bf \theta})\pi({\bf \theta}|{\bf \Lambda},K)},
\end{equation}
\begin{equation} 
Z(d_i|{\bf \Lambda})=\int{{\rm d}{\bf \theta}\mathcal{L}(d_i|{\bf \theta})\pi({\bf \theta}|{\bf \Lambda})},
\end{equation}
where $\pi({\bf \theta}|{\bf \Lambda},K)$ is the joint mass-spin distribution for the $K$ channel under the hyper-parameter ${\bf \Lambda}$, and $\pi({\bf \theta}|{\bf \Lambda}) = \sum_{K}r_K\pi({\bf \theta}|{\bf \Lambda},K)$. Therefore, one can find that $\sum_{K}{P_{K}^{i}}=1$, i.e., $P_{K}^{i}$ is normalized. We calculate the probabilities of formation channels for the 69 events and summarize them in Tab.~\ref{tab:pchannel} (the events that are most likely containing 2G BHs are highlighted in bold face). 
 
\clearpage

\begin{longtable}{lccccc}
\caption{Probabilities of formation channels for events in GWTC-3}\label{tab:pchannel}
\\
\toprule
\hline
Events   & Field  & 1G+1G & 1G+2G & 2G+1G & 2G+2G \\ 
\midrule
\endfirsthead

\multicolumn{6}{c}{\autoref{tab:pchannel} ({\it Continued})}\\
\toprule
\hline
Events   & Field  & 1G+1G & 1G+2G & 2G+1G & 2G+2G \\ 
\midrule
\endhead
\bottomrule
\\
\multicolumn{6}{c}{\autoref{tab:pchannel} ({\it Continued on next page})}\\
\endfoot
\bottomrule
\\
\endlastfoot
GW150914\_095045&0.011&0.974&0.004&0.009&0.0 \\
GW151012\_095443&0.532&0.467&0.0&0.001&0.0 \\
GW151226\_033853&0.971&0.029&0.0&0.0&0.0 \\
GW170104\_101158&0.095&0.866&0.001&0.002&0.0 \\
GW170608\_020116&0.986&0.014&0.0&0.0&0.0 \\
\textbf{GW170729\_185629}&0.001&0.09&0.011&0.868&0.032 \\
GW170809\_082821&0.036&0.958&0.002&0.005&0.0 \\
GW170814\_103043&0.045&0.949&0.002&0.005&0.0 \\
GW170818\_022509&0.017&0.956&0.005&0.013&0.0 \\
GW170823\_131358&0.013&0.964&0.005&0.016&0.0 \\
GW190408\_181802&0.214&0.746&0.001&0.001&0.0 \\
GW190412\_053044&0.784&0.218&0.0&0.0&0.0 \\
GW190413\_134308&0.005&0.715&0.008&0.238&0.004 \\
GW190421\_213856&0.008&0.942&0.008&0.037&0.001 \\
GW190503\_185404&0.008&0.952&0.004&0.029&0.0 \\
GW190512\_180714&0.637&0.356&0.0&0.0&0.0 \\
GW190513\_205428&0.106&0.889&0.002&0.015&0.0 \\
\textbf{GW190517\_055101}&0.003&0.065&0.019&0.9&0.016 \\
\textbf{GW190519\_153544}&0.0&0.0&0.0&0.728&0.272 \\
\textbf{GW190521\_030229}&0.0&0.0&0.0&0.0&1.0 \\
GW190521\_074359&0.005&0.909&0.004&0.068&0.001 \\
GW190527\_092055&0.053&0.944&0.002&0.008&0.0 \\
\textbf{GW190602\_175927}&0.0&0.0&0.0&0.319&0.681 \\
\textbf{GW190620\_030421}&0.0&0.03&0.003&0.86&0.106 \\
GW190630\_185205&0.033&0.964&0.003&0.003&0.0 \\
\textbf{GW190701\_203306}&0.0&0.001&0.0&0.871&0.128 \\
\textbf{GW190706\_222641}&0.0&0.0&0.0&0.618&0.382 \\
GW190707\_093326&0.984&0.015&0.0&0.0&0.0 \\
GW190708\_232457&0.771&0.228&0.0&0.0&0.0 \\
GW190720\_000836&0.975&0.025&0.0&0.0&0.0 \\
GW190727\_060333&0.014&0.945&0.009&0.03&0.001 \\
GW190728\_064510&0.975&0.025&0.0&0.0&0.0 \\
GW190803\_022701&0.016&0.941&0.005&0.014&0.0 \\
GW190828\_063405&0.036&0.959&0.003&0.008&0.0 \\
GW190828\_065509&0.765&0.232&0.0&0.0&0.0 \\
GW190910\_112807&0.005&0.859&0.006&0.071&0.001 \\
GW190915\_235702&0.035&0.946&0.003&0.011&0.0 \\
GW190924\_021846&0.998&0.002&0.0&0.0&0.0 \\
GW190925\_232845&0.46&0.552&0.0&0.0&0.0 \\
\textbf{GW190929\_012149}&0.006&0.401&0.0&0.572&0.016 \\
GW190930\_133541&0.981&0.019&0.0&0.0&0.0 \\
GW190413\_052954&0.036&0.952&0.003&0.007&0.0 \\
GW190719\_215514&0.095&0.887&0.005&0.025&0.001 \\
GW190725\_174728&0.994&0.006&0.0&0.0&0.0 \\
GW190731\_140936&0.017&0.94&0.007&0.037&0.001 \\
\textbf{GW190805\_211137}&0.009&0.489&0.026&0.432&0.026 \\
GW191105\_143521&0.988&0.012&0.0&0.0&0.0 \\
\textbf{GW191109\_010717}&0.0&0.0&0.0&0.261&0.739 \\
GW191127\_050227&0.012&0.825&0.014&0.159&0.01 \\
GW191129\_134029&0.992&0.008&0.0&0.0&0.0 \\
GW191204\_171526&0.977&0.023&0.0&0.0&0.0 \\
GW191215\_223052&0.267&0.731&0.001&0.002&0.0 \\
GW191216\_213338&0.981&0.019&0.0&0.0&0.0 \\
GW191222\_033537&0.006&0.943&0.007&0.038&0.002 \\
GW191230\_180458&0.002&0.507&0.009&0.431&0.03 \\
GW200112\_155838&0.015&0.974&0.004&0.006&0.0 \\
GW200128\_022011&0.011&0.908&0.009&0.063&0.003 \\
GW200129\_065458&0.015&0.977&0.004&0.007&0.0 \\
GW200202\_154313&0.99&0.01&0.0&0.0&0.0 \\
GW200208\_130117&0.013&0.975&0.004&0.009&0.0 \\
GW200209\_085452&0.023&0.962&0.004&0.014&0.0 \\
GW200219\_094415&0.016&0.947&0.005&0.019&0.0 \\
GW200224\_222234&0.007&0.946&0.01&0.04&0.001 \\
GW200225\_060421&0.59&0.382&0.0&0.001&0.0 \\
GW200302\_015811&0.054&0.923&0.004&0.013&0.0 \\
GW200311\_115853&0.016&0.973&0.003&0.005&0.0 \\
GW200316\_215756&0.975&0.025&0.0&0.0&0.0 \\
GW191103\_012549&0.981&0.019&0.0&0.0&0.0 \\
GW200216\_220804&0.008&0.828&0.007&0.173&0.01 \\

\end{longtable}

\end{document}